\documentclass[11pt, reprint]{revtex4-1}

\usepackage{graphics}
\usepackage{amssymb}
\usepackage{amsmath}
\usepackage{pstricks}
\usepackage{float}

\renewcommand{\tensor}[1]{\mathsf{#1}}
\renewcommand{\vec}[1]{\mathbf{#1}}

\begin{document}
\title{Four-component united-atom model of bitumen}

\author{J. S. Hansen}
\email{jschmidt@ruc.dk}
\affiliation{
  DNRF Centre ``Glass and Time'', IMFUFA, 
  Department of Science, Systems and Models, 
  Roskilde University, Postbox 260, 
  DK-4000 Roskilde, Denmark
}

\author{Claire A. Lemarchand}
\affiliation{
  DNRF Centre ``Glass and Time'', IMFUFA, 
  Department of Science, Systems and Models, 
  Roskilde University, Postbox 260, 
  DK-4000 Roskilde, Denmark $\mbox{}$
}

\author{Erik Nielsen}
\affiliation{
  Danish Road Directorate, 
  Guldalderen 12, 
  DK-2640 Hedehusene, Denmark
}

\author{Jeppe C. Dyre}
\affiliation{
  DNRF Centre ``Glass and Time'', IMFUFA, 
  Department of Science, Systems and Models, 
  Roskilde University, Postbox 260, 
  DK-4000 Roskilde, Denmark $\mbox{} $ $\mbox{}$ 
}

\author{Thomas Schr{\o}der}
\affiliation{
  DNRF Centre ``Glass and Time'', IMFUFA, 
  Department of Science, Systems and Models, 
  Roskilde University, Postbox 260, 
  DK-4000 Roskilde, Denmark $\mbox{} $ $\mbox{} \mbox{}$ 
}

\begin{abstract}
We propose a four-component molecular model of
bitumen. The model includes realistic chemical constituents and 
introduces a coarse graining level that suppresses the highest frequency
modes. Molecular dynamics simulations of the model are be carried out
using Graphic-Processor-Units based software in time spans
in order of microseconds, and this enables the study of slow relaxation
processes characterizing bitumen.  
This paper focuses on the high-temperature dynamics 
as expressed through the mean-square displacement, 
the stress autocorrelation function,  and rotational relaxation. The
diffusivity of the individual molecules changes little as a function
of temperature and reveals distinct dynamical time scales as a result
of the different constituents in the system. Different
time scales are also observed for the rotational relaxation. 
The stress autocorrelation function features a slow 
non-exponential decay for all temperatures studied. From the
stress autocorrelation function, the shear viscosity and shear 
modulus are evaluated at the
highest temperature, showing a viscous response at 
frequencies below 100 MHz. The model predictions of viscosity and
diffusivities are compared to experimental data, giving reasonable agreement. 
The model shows that the asphaltene, resin and resinous oil tend to
form nano-aggregates. The characteristic dynamical 
relaxation time of these aggregates is different from the
homogeneously distributed parts of the system, leading to strong
dynamical heterogeneity. 
\end{abstract}

\maketitle

\section{Introduction}
Refined bitumen (or asphalt) is a residual product of the refinery process of
crude oil, and is highly viscous at room temperatures.
The chemical composition of bitumen is very complex in that it is composed of 
up to 10$^6$ relatively large
molecules \cite{wiehe:1996,rogel:ef:2003,artok:ef:1999,murgich:ef:1996}
with molar masses above 200 g/mol. Bitumen is not chemically unique in the sense
that its composition depends on the crude oil source,  the age of the
bitumen \cite{barth:1962}, and possible chemical
modifiers~\cite{zhang:ef:2008}. The functional chemistry includes
heteroatoms like sulfur and nitrogen, cyclic structures, aromatics
compounds and saturated hydrocarbons~\cite{barth:1962}, and these
enter larger and complicated molecular structures. A correct and
detailed chemical categorization is not feasible, and bitumen is
often characterized simply by its rheological properties such as bulk
and shear moduli  and penetration depth~\cite{reed:2003}. 
A fundamental understanding on the molecular level is, however, 
necessary if one wishes to understand and control the mechanical 
properties. 

The aim of the present work is to do just that, i.e., we wish to study
dynamical properties of bitumen through molecular modelling. 
A further impetus of this work
lies on the definition and characterization of a ``Cooee-bitumen
model'' with the aim to bring new insights on how to reduce the
rolling resistance generated between the tyre and the
pavement~\cite{schmidt:2012}. Molecular simulations has previously 
been used to investigate bitumen properties,  for example, the density and
structure in asphaltene systems~\cite{rogel:ef:2003} and molecular
alignment~\cite{murgich:ef:1996}. 
Recently, Zhang and Greenfield (Z \& G) published a series of
papers
\cite{zhang:ef:2007,zhang:ef:2007:2,zhang:jcp:2007,zhang:ef:2008,zhang:jcp:2010}, 
wherein they study structural, thermodynamical and dynamical
properties  of different bitumen models with and without polymer
additives. To this end the authors used atomistic molecular
dynamics simulations. Their bitumen models  
are based on a tertiary mixture where each chemical component represents a
distinct constituent : high-weight 
asphaltenes, aromatic maltenes (resin) and 
saturated maltenes ($n$-alkanes).  In their simulations they applied an
all-atom force  field that includes the interactions between 
both hydrogen and carbon and 
hydrogen and sulfur, as well as the hydrogen-carbon bonds. 
While this allows for standard and well tested force field 
parametrizations and gives very detailed description of the system, it
also includes very fast modes which are unlikely to couple to the long
relaxation times characterizing viscous liquids like bitumen.  
The inclusion of these fast modes necessitates a 
small integration time step in the numerical intregration of Newton's
equations of motion, and Z \& G were 
able to simulate a time span in the order of 10 nsec., even
using parallelized molecular dynamics software~\cite{zhang:jcp:2010}. 
For the aromatic maltene constituent Z \& G used dimethylnaphtalene 
or closely related molecular structures. 
This relatively low weight molecule has a boiling 
point of around 278 $^{\mathrm{o}}$C ~\cite{lide:1991}, 
which is much lower than that of the other
constituents and dimethylnaphtalene is therefore likely to be
removed from the crude oil during the refinery process at approximately 500
$^{\mathrm{o}}$C, see Ref. \onlinecite{barth:1962}. 

We propose a new four-component bitumen model, based on the
Hubbard-Stanfield classification~\cite{hubbard:1948,rostler:1965} 
The model is coarse grained such that the hydrogen
atoms are not explicitly included, but only implicitly 
through parametrization of the particle interactions. This allows for 
larger time steps and, more importantly, it reduces significantly the
number of bond-, angle- and pair-forces one needs to evaluate 
in each time step. The molecular dynamics 
simulations are carried out using the Graphical-Processor-Unit (GPU) 
based molecular dynamics software package RUMD~\cite{bailey:2012}. In this way
time spans in the order of $\mu$sec. are accessible using a single
graphics card. We are thus able to study much slower relaxation
phenomena in bitumen compared to what was previously possible.

In this paper we focus on the high-temperature dynamical properties of
the model and identify the temperature where the relaxation time
becomes very long. Specifically, we investigate
characteristic relaxation times for collective, chemical and single
molecule quantities. From this we show that the dynamical
properties of the model are quite different from those of Z \& G's model, in
that the dynamics are much slower and changes only little in the high
temperature regime. For relatively low temperatures, the dynamics is
characterized by complex glass-like behavior, where the relaxation
times widely exceed the accessible simulation times.

The paper is organized as follows. Section~\ref{sec:model} describes
the parameters chosen in the model and the simulation method. Section~\ref{sec:results}
contains results and discussion of diffusivity, viscous properties and
rotational dynamics of the model bitumen. Finally, a summary of the
paper is provided in section~\ref{sec:summary}.

\section{Model and parametrization \label{sec:model}}
As stated in the introduction, a detailed chemical model for bitumen
is not possible. We do therefore not target any specific bitumen,
but aim at capturing the characteristic properties of a typical bitumen. 
We conjecture that an explicit inclusion of the hydrogen 
atoms is not important for our modelling purpose, even if it may be useful for
other specific purposes (see for example Ref.~\onlinecite{toxvaerd:1990}). We
therefore coarse grain the chemical structures which enables us to
investigate longer time dynamics. 

\subsection{Parametrization}
\begin{figure}[h]
  \scalebox{0.4}{\includegraphics{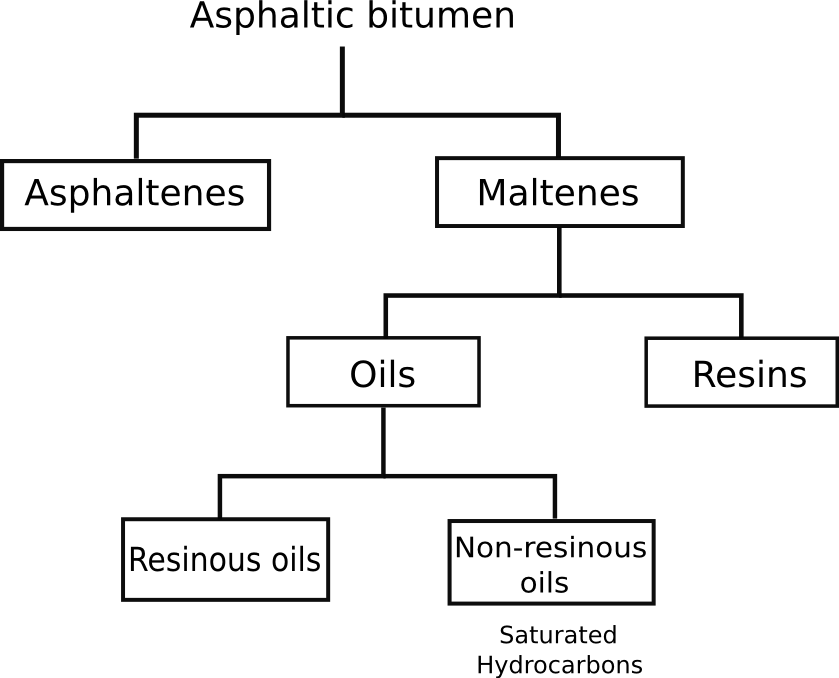}}
  \caption{
    \label{fig:bdiagram}
    Classification scheme of bitumen by Hubbard and Stanfield. From
    Ref. \onlinecite{rostler:1965}. 
  }
\end{figure}

The bitumen constituents are characterized using different separation
techniques \cite{rostler:1959,barth:1962,reed:2003}. We will here use
a general classification scheme by Hubbard and Stanfield
\cite{hubbard:1948,rostler:1965}, see Fig. \ref{fig:bdiagram}. We let unsaturated
hydrocarbons and the aromatic/cyclic compounds without any heteroatoms
like sulfur fall into the catagory resinous
oils~\cite{barth:1962}. Also, we interpret the class non-resinous oils 
to be paraffin wax, i.e. staturated hydrocarbons. This motivates a
four-constituent model composed of (i) asphaltene, (ii) resin, (iii) 
resinous oil and (iv) saturated hydrocarbon. 
The model also follow the SARA classification
(Saturates-Aromatics-Resins-Asphaltenes)\cite{astm:2007}; although
the unsaturated hydrocarbons and the aromatic/cyclic compounds fall
under the catagory 'Aromatics' in the SARA classification, we prefer
to use the term resinous oil to follow Hubbard and Stanfield.   

\begin{figure}[ht]
  \scalebox{0.1}{\includegraphics{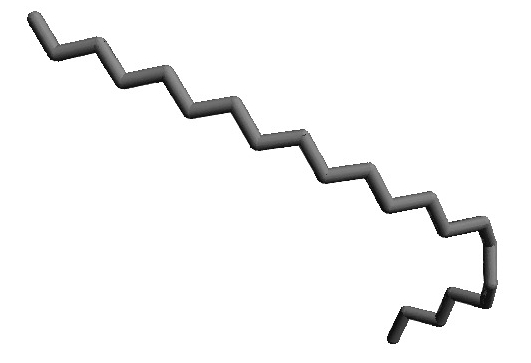}}
  \scalebox{0.17}{\includegraphics{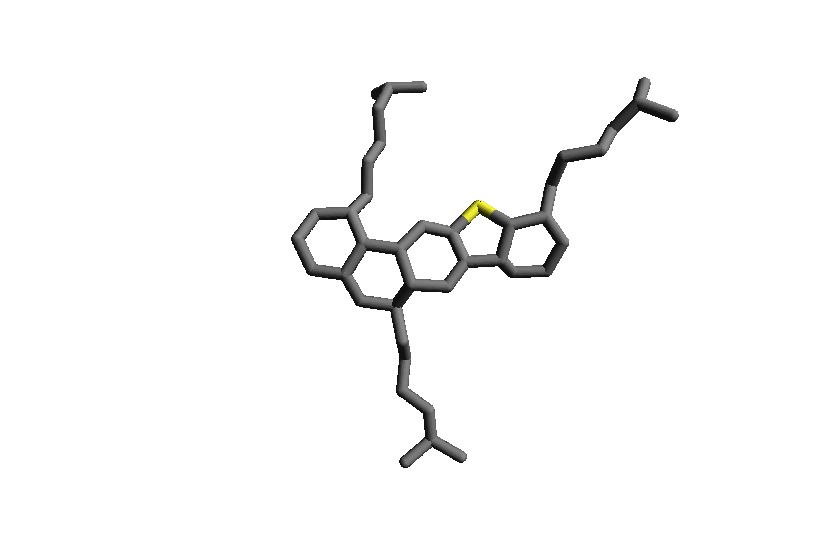}} \\
  \scalebox{0.20}{\includegraphics{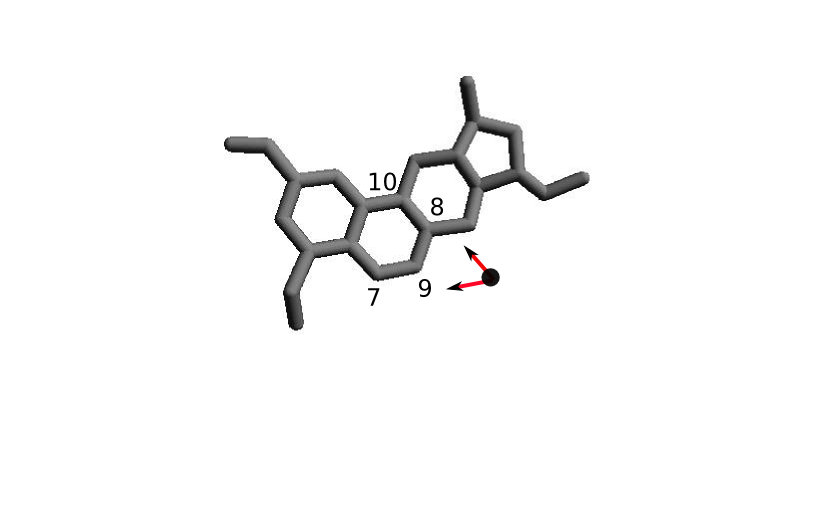}}
  \scalebox{0.20}{\includegraphics{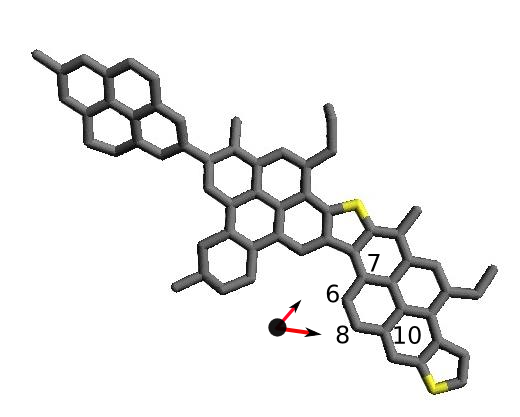}}
  \caption{\label{fig:molecules}
    (Color online). United-atom-unit models of the 
    constituent molecules in the bitumen (or asphalt) model. 
    Top left: Docosane. Top right: Resin.
    Lower left: Resinous oil. Lower right: Asphaltene.
    Yellow indicates sulfur atoms. Numbers and arrows indentify
    bond-vectors used in the data analysis, see text for details}
\end{figure}

A single molecule is used to represent each constituent.
Based on NMR studies Artok~\cite{artok:ef:1999} proposed an asphaltene
structure, shown in Fig. \ref{fig:molecules}, which is adopted here. This
structure was also used by Z \& G which they call asphaltene 1. 
For the resinous oil and resin, we use the structures by Rossini et 
al.~\cite{rossini:1953} and Murgich et al.~\cite{murgich:ef:1996},
respectively. In Ref. \onlinecite{zhang:ef:2007}, Z \& G argued that
the saturated hydrocarbons can be modelled by 
docosane ($n$-C$_{22}$), since this molecule represents the average
chain length found by Storm et al.~\cite{storm:ef:1994}. It is also
used here. All four molecular structures are shown in Fig. \ref{fig:molecules}. 

We represent the methyl (CH$_3$),
methylene (CH$_2$), and methine (CH) groups  by the same Lennard-Jones
particle. This is then a united atomic unit (UAU)  with mass 13.3 g/mol and 
Lennard-Jones parameters $\sigma$ = 3.75 {\AA} and $\epsilon = 75.4$ 
$k_BT$ are estimated from the OPLS force field~\cite{jorgensen:jacs:1988}.
The sulfur atoms are represented by the
same Lennard-Jones interaction, but with a mass of 32 g/mol. 

The force field is given through the total potential energy of the
system 
\begin{equation}
\begin{aligned}
  U(\vec{r})&=   \sum_{i} \sum_{j>i} 4\epsilon\left[
          \left(\frac{\sigma}{r_{ij}}\right)^{12} -
          \left(\frac{\sigma}{r_{ij}}\right)^{6}
          \right] \\
          &+\frac{1}{2}\sum_{\text{\tiny{bonds}}}k_s(r_{ij}-l_{\text{\tiny{b}}})^2\\
  &+\frac{1}{2}\sum_{\text{\tiny{angles}}}k_{\theta}(\cos \theta - \cos
  \theta_0)^2\\
  &+\sum_{\text{\tiny{dihedrals}}}\sum_{n=0}^5 c_n\cos^n \phi.
\end{aligned}
\label{eq:forcefield}
\end{equation}
The force field thus includes pair interactions as well as interactions due to 
(flexible) bonds, angles and dihedral angles. The bonds, angles and
dihedral angles are specified via bonds between the UAUs. 
UAUs belonging to the same bond, angle and
dihedral angle do not interact via the pair interaction. From the force field it
can bee seen that the model has a large 
parameter space. To reduce the number of free parameters, we
assume that all bonds have the same length $l_b=1.46$ {\AA}. 
This particular value of $l_b$ is chosen to be the average value of all the
bonds in the system ($1.46 \pm 0.07$ {\AA}) computed from energy
minimization of the individual molecules~\cite{avogadro}. This is, of
course, a simplification since energy minimization of, for
example, the asphaltene gives the single bonded units a bond length of 
around 1.50-1.54 {\AA} and the double
bonded units a bond length 1.38-1.41 {\AA}. These values again 
depend on the specific molecule and amount to very many different bond
length, so we choose an average value for simplicity. 
The same argument lies behind the
choices of the angles and dihedral angles, but we here use
two distinct angles and three distinct dihedral angles.   
We use the Generalized Amber Force Field (GAFF)~\cite{wang:jcc:2004} 
parametrization for bond and angle parameters,
see Table \ref{table:params}. For the saturated hydrocarbons and
the side chains in the asphaltene and resin, we
use the Ryckaert-Bellemans parameters~\cite{catlow:1989}. These
parameters are optimized with respect to liquid butane, but they
capture the important \textit{cis}, \textit{trans} and \textit{gauche}
configurations reasonably well. We are thereby left with a single free
parameter, namely the zero-force coefficient $c_1$ for dihedral angles 
$\phi=180$ and 0 degrees in the ring structures.  

\begin{figure}
  \scalebox{0.25}{\includegraphics{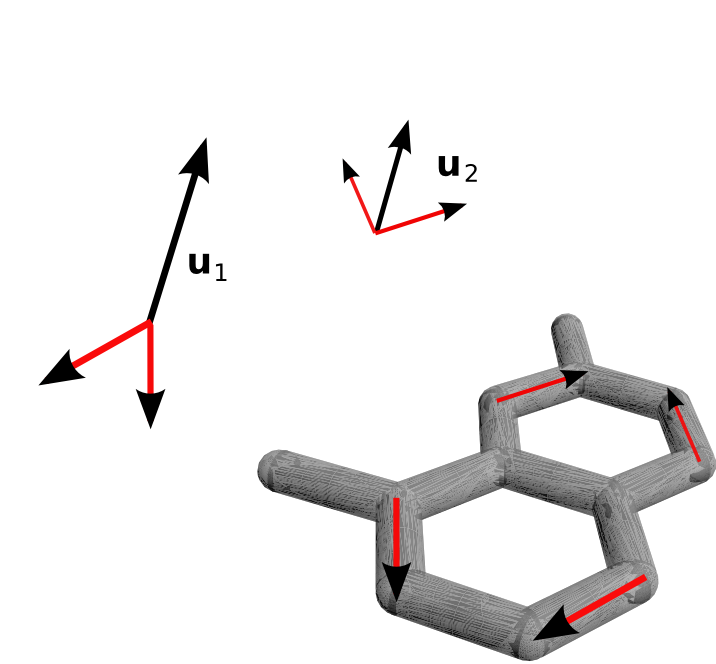}}
  \scalebox{0.30}{\includegraphics{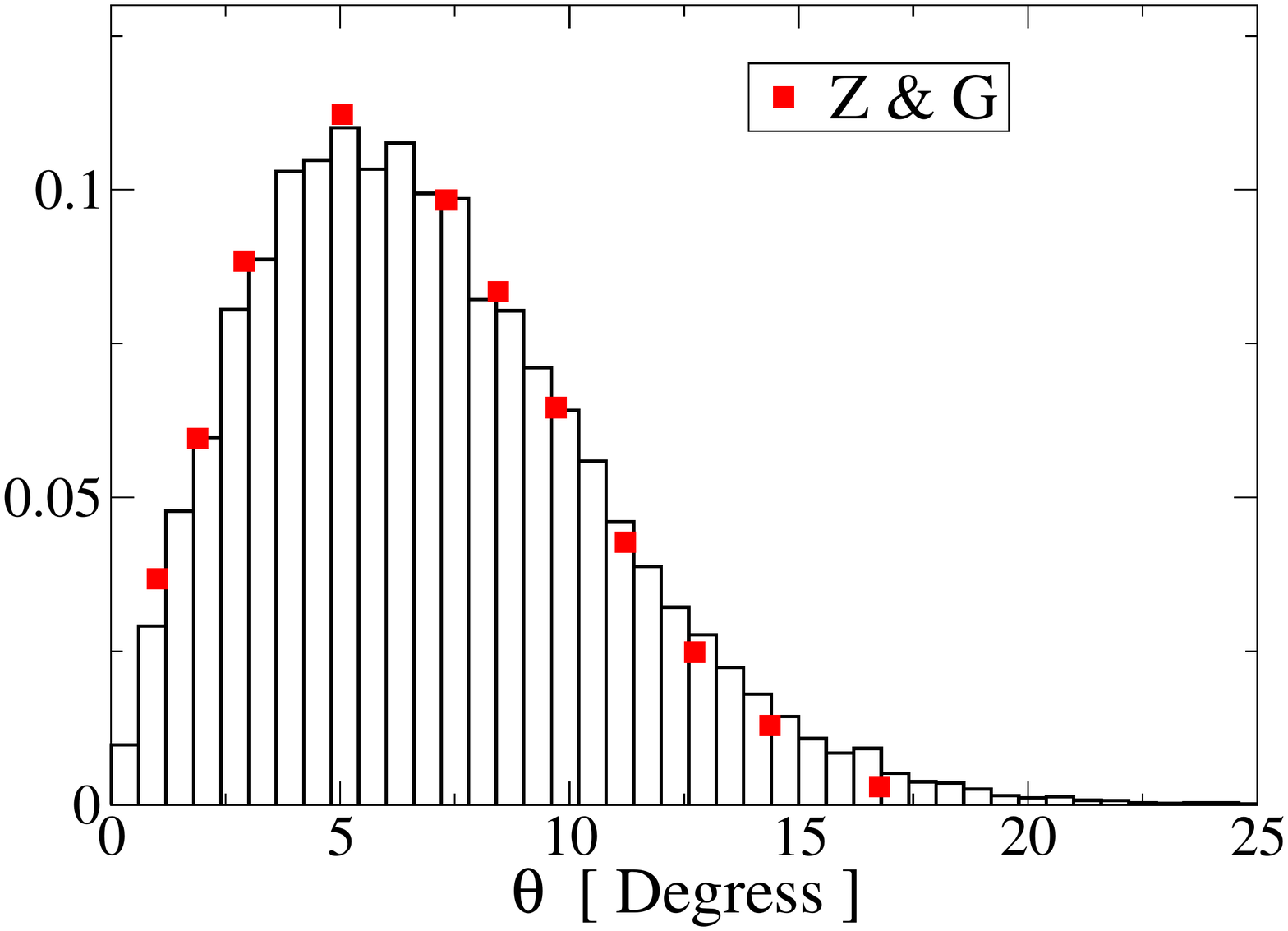}}
  \caption{\label{fig:alignment}
    (Color online). 
    Left: Definition of the alignment angle, 
    $\theta$, in 1,7-dimethylnaphtalene.
    Right: Angle distribution for pure 1,7-dimethylnaphtalene at the state
    point $\rho$=994 kg/m$^3$ and $T=358.15$ K. Reference data are taken from
    Ref. \onlinecite{zhang:ef:2007:2}.  
  }
\end{figure}

The $c_1$ parameter is estimated from one structural criterion, namely 
the alignment between the aromatic rings in a one-component 
system of 1,7-dimethylnaphtalene (also used by Z \& G). 
This particular criterion is chosen because of the simple molecular 
structure of 1,7-dimethylnaphtalene. Let the vector $\vec{u}_1$
be defined as the cross product $\vec{u}_1=\vec{r}_1\times\vec{r}_2$,  
where $\vec{r}_1$ and $\vec{r}_2$ are two vectors parallel with two bonds
in one of the ring structure of the molecule, as shown in Fig.~\ref{fig:alignment}.
Thus $\vec{u}_1$ is a vector normal to the ring structure. Likewise, the vector
$\vec{u}_2$ is defined as normal to the second ring structure. The
angle between $\vec{u}_1$ and $\vec{u}_2$ then defines the ring alignment
angle, $\theta = \cos^{-1}(\vec{u}_1\cdot\vec{u}_2/|\vec{u}_1||\vec{u}_2|)$. 
Figure~\ref{fig:alignment} shows the histogram 
of the alignment angle for 1,7-dimethylnaphtalene at mass density 
994 kg/m$^3$ and temperature $T=358.15$ K. The values
$c_1 = 20$ kcal/mol and $c_1 = -20$ kcal/mol were chosen for the $\phi=180$ 
and $\phi = 0$ degrees zero-force dihedral angles, respectively, as 
they lead to a good agreement 
between the all-atom model used by Z \& G and the united-atom model
used in the present work.  Table~\ref{table:params} lists all
parameters used in the model.
\begin{table} [h]      
  \begin{tabular}{lcc}
    \hline \hline
    Bonds                   & $l_b$  & $k_s$ 
    \\
    & [\AA]   & [kcal/mol]
    \\	
    \hline
    All & 1.46     & 403 \\
    \hline \hline
    Angles                        & $\theta_0$      & $k_\theta$  \\
    & [Degrees]  & [kcal/(mol rad$^2$)]  \\
    \hline
    Aromatic/cyclo            &    120      & 108 \\
    Aliphatic           &    106      & 70 \\
    \hline \hline
    Dihedrals                   & Angle       &  $c_n$   \\            
    &  [Degrees]  &   [kcal/mol] \\
    \hline
    Aromatic/cyclo  & 180  &  20  (n=1)\\ 
    Aromatic/cyclo  &   0  & -20  (n=1)\\
    Linear/aliphatic &   &  Ryckaert-Belleman \\
    \hline \hline
  \end{tabular}
  \caption{\label{table:params} 
    Force-field parameters, see Eq. (\ref{eq:forcefield}).
  }
\end{table}

\subsection{Simulation Method}
In all simulations we used $N_A=10$ asphaltene, $N_R=10$ resin,
$N_{RO}=10$ resinous oil and $N_D=82$ docosane molecules, giving a
total of 3114 UAUs. The mass fraction corresponds closely to Z \& G's
mixture 1~\cite{zhang:ef:2007},  
for which dimethylnaphtalene represents both the resin and resinous oil, but
we used twice as many molecules. Initially, the molecules
center-of-mass are arranged on a simple low-density lattice and the
system is compressed (while integrating the equation of
motion) to the desired density. We use a target density of
$\rho=994$ kg/m$^3$, and the quench is carried out for a 
temperature of $T=603$ K. The system was coupled to a
Nos\'{e}-Hoover thermostat~\cite{nose:mp:1984,hoover:pra:1985} 
and the equations of motion were
integrated using a leap-frog integrator
scheme~\cite{frenkel:1996}. After the compression the volume is kept
fixed in order to perform NVT ensemble simulations. A first production run is
carried out at $T=603$ K over a total time span of 0.85
$\mu$sec. During this run the system reaches equilibrium
(see the discussion below). The final configuration from this state
point is used to reduce the temperature from
$T=603$ K to $T=528, 490$ and 452 K at a rate of
2.2 K per nanosecond. In this way we study the systems at four
different temperatures. For $T \leq 528$ K production runs of 1.7
$\mu$sec. were performed. For all temperatures the production runs
were divided into 10 intervals of same time span (0.085  $\mu$sec.
and 0.17 $\mu$sec. for $T$ = 603 K and $T \leq$ 528 K,
respectively), which enables us to follow the equilibration of the
system. We only use the last eight sample intervals for data analysis
in order to avoid the initial fast relaxations in the system. 
 
In order to highlight the different dynamics we also studied 
the system at a relatively low temperature and far from equilibrium. To this end a
dilute and high-temperature system is rapidly quenched to the 
target density and $T=301$ K. This system is also simulated over a
time span of 1.7 $\mu$sec. 

To determine the integration time step, the energy drift was  
investigated by performing a series of NVE simulations of the bitumen mixture at 
the desired density and for different temperatures over one million
time steps. For low temperatures and time steps larger than 3.4
femtoseconds, we noticed a slight systematic 
increase in the energy fluctuations (quantified by the energy standard
error). To be conservative, we chose a time step of 1.7 femtoseconds
for $T\leq$528 K and 0.85 
femtoseconds for T=603 $K$. These time steps are around 35-70 \%
larger than those used by Z \& G~\cite{zhang:jcp:2007}, and are made possible due
to the coarse graining. This enables us to simulate a time span
of up to 1.7 $\mu$sec. using one billion time steps. 

%
%

\section{Results and discussion}
\label{sec:results}
This study analyzes (i) the average diffusivity of the four different 
chemical components, (ii) the collective viscous properties through the
stress (or equivalently the pressure tensor) autocorrelation
function and (iii) the orientational dynamics of single resinous oil  
and asphaltene molecules.

\subsection{Diffusivity\label{sec:msd}}
\begin{figure}
  \scalebox{0.275}{\includegraphics{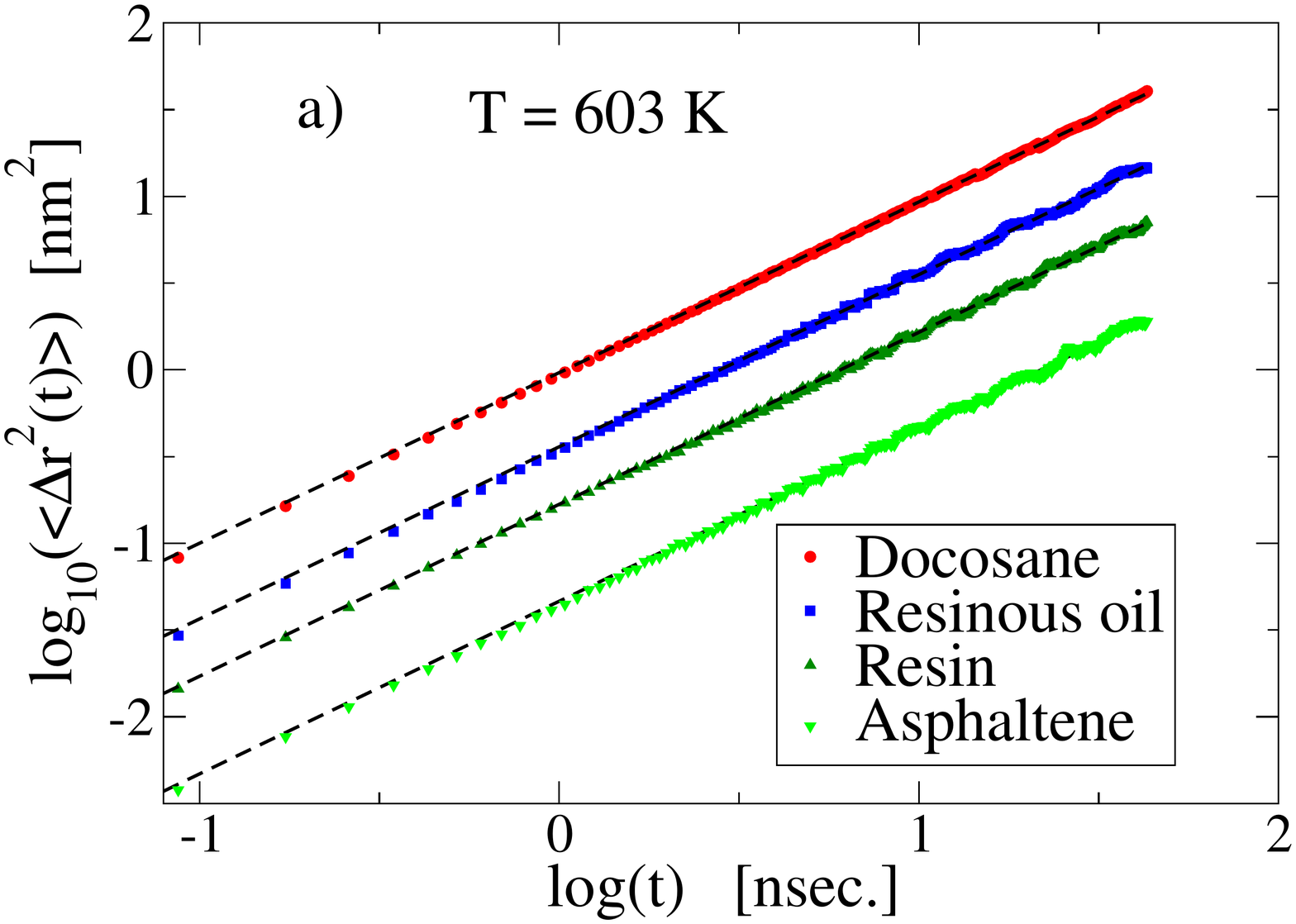}}
  \scalebox{0.275}{\includegraphics{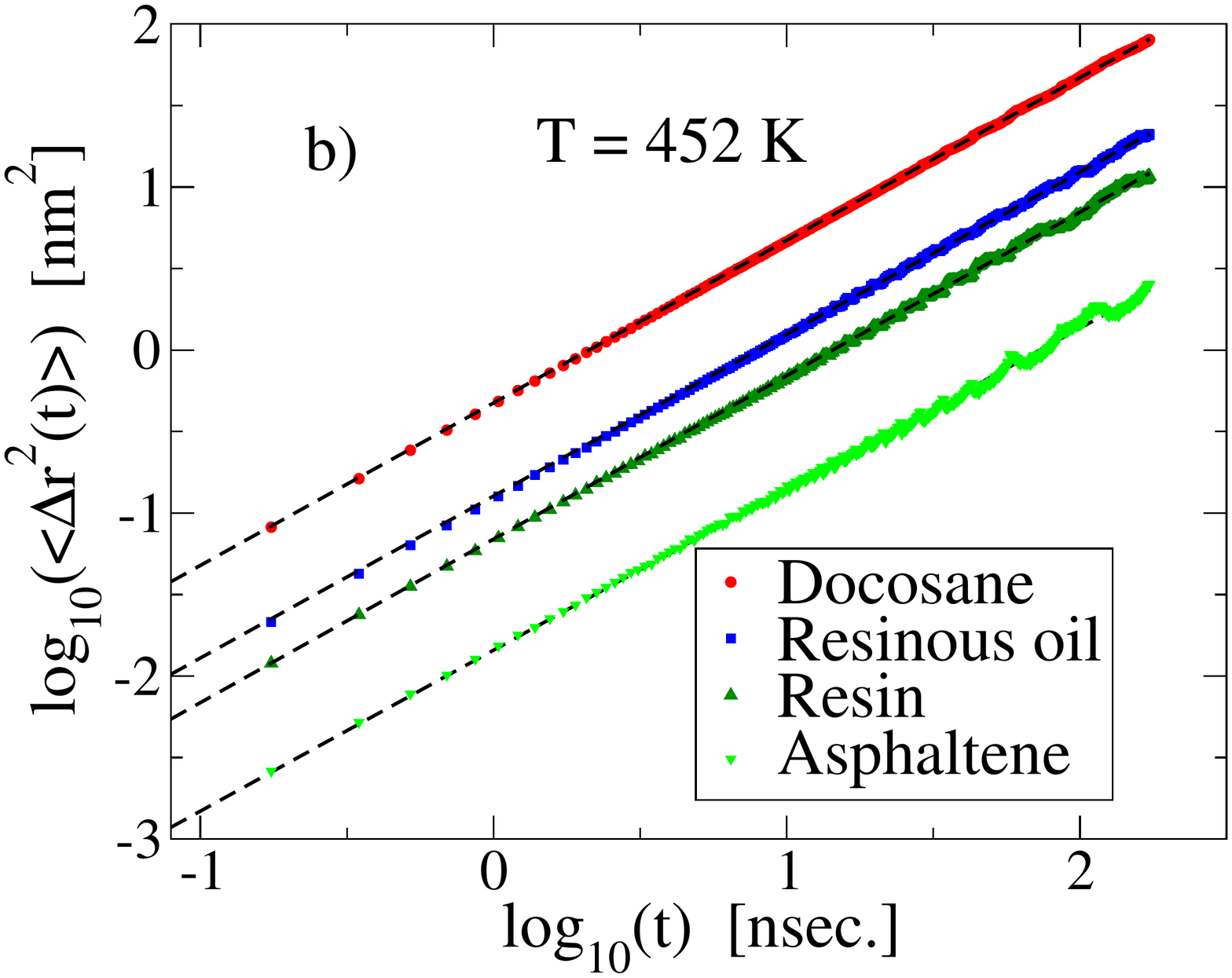}}
  \caption{\label{fig:msd}
    (Color online). Mean-square displacement of the molecules. 
    Dashed lines are the best fits of the form given in Eq. (\ref{eq:loglog})
    to the data points, for $t$ smaller than 1.7 nsec.
    (a): $T$=603 K, (b): $T$ = 452 K. Data are averaged over the 8 
    sample intervals, $t=85$ nsec. and 170 nsec. for $T$=603 K and
    $T$=452 K.    
  }
\end{figure}

\begin{figure}
  \scalebox{0.275}{\includegraphics{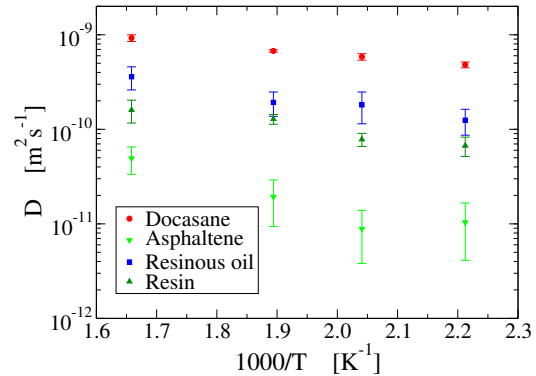}}
  \caption{\label{fig:arrh_diff}
    Arrhenius plot for the four chemical constituents in the bitumen
    model. The error bars are standard deviations
    estimated from the eight last sample intervals of each production
    run.     
  }
\end{figure}

The diffusivity of the molecules is evaluated through the mean-square
displacement of the center-of-mass of each molecule. If the
center-of-mass position of molecule $i$ 
is denoted $\vec{r}_i$, the mean-square displacement of molecules of a
certain type is defined as \cite{hansen:2006}: 
$ \langle \Delta \vec{r}^2 (t)\rangle = \frac{1}{N}\sum_i^{N}
\Delta \vec{r}_i^2 (t)$, where index $i$ runs over molecules of
that type. In the normal diffusive regime  
\begin{eqnarray}
\log_{10}\left[\langle \Delta \vec{r}^2 (t)\rangle \right] = 
\log_{10}(6 D)  + \log_{10}(t) ,
\label{eq:loglog}
\end{eqnarray}
where $D$ is the diffusion coefficient. In Fig. \ref{fig:msd},
the mean square displacements of the four
components is plotted for 452 K and 603 K.  
It is seen that the molecules are in the normal diffusive regime at
both temperatures, for times longer than 1 nanosecond. The corresponding diffusion coefficients
are plotted in Fig. \ref{fig:arrh_diff} for different temperatures. 
Upreti and Mehrotra~\cite{upreti:2002} reported diffusivities in the order of
0.1 - 1.0 $\times 10^{-9}$ m$^2$/s for small weight gases in bitumen over a 
temperature range of 20-200 $^0$C, which is in reasonable agreement
with our results in view of the fact that we study higher weight molecules. 
From Fig. \ref{fig:arrh_diff} it follows that $D_D > D_{R0} >
D_{R} > D_A$, where $D_D$ is the diffusivity for docosane, $D_{R0}$
for resinous oil, $D_R$ for resin and $D_A$ for asphaltene. This, of
course, agrees with the fact that the mobility decreases with
increasing molecular size.  
 
Both asphaltene and docosane have diffusion coefficients that are
lower than the values reported by Z \& G~\cite{zhang:jcp:2007}, who found
$D_{D} = 1\times 10^{-9}$ m$^2$/s and $D_{A}=2.5 \times
10^{-10}$ m$^2$/s at $T=440.5$ K. This discrepancy is
likely due to the current model replacing 1,7-dimethylnaptalene with
the resin and resinous oil compounds. These relatively large
structures hinder the motion of the compounds significantly
compared to the smaller 1,7-dimethylnaphtalene. This may also explain
another difference between the two models: we find a
large difference in diffusivity between the docosane and asphaltene
($D_D \approx 20 D_A$) whereas Z \& G's model shows a
difference of only a factor of 5. 

\begin{figure}
  \scalebox{0.275}{\includegraphics{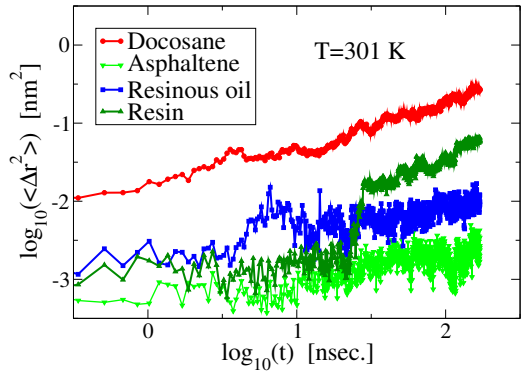}}
  \caption{\label{fig:msd_T40}
    (Color online). Mean-square displacement of the molecules for the
    quickly quenched system. Additional averaging over time lags was
    not carried out as in Fig.~\ref{fig:msd}.
  }
\end{figure}

The mean-square displacement for a system rapidly quenched to the
temperature $301$ K was also 
studied and is plotted in Fig. \ref{fig:msd_T40}. In order to highlight
the details and as opposed to what was done in Fig.~\ref{fig:msd}, no
sample averaging is done here. 
The docosane molecules still exhibit an approximate diffusive
motion whereas the larger resin and resinous oil molecules perform
vibrational motion  around a given position (arrested), and then on
average make a rapid movement (jump). The resin molecules eventually
enter an approximately diffusive regime. The arrest and jump motions are both random
events, which are also observed for simple models of
glasses~\cite{berthier:rmp:2011,edinger:jcp:2012}. 
This behavior shows the inherent heterogeneous dynamics present in the
bitumen model where the asphaltenes are immobile, on the time scales
shown in the figure, whereas the docosanes approaches normal
diffusivity. 

\subsection{Viscous properties}
To study the viscosity we use the Irving-Kirkwood~\cite{irving:1951} expression
for the pressure tensor 
\begin{eqnarray}
\tensor{P} = \frac{1}{V}\sum_{i} \left[
\frac{\vec{p}_i\vec{p}_i}{m_i} + \sum_{j>i}\vec{r}_{ij}
\vec{F}_{ij}\right] \ ,
\end{eqnarray}
where $\vec{p}_i$ is the momentum of molecule $i$, $m_i$ its mass,
$\vec{r}_{ij}= \vec{r}_i-\vec{r}_j$ is the center-of-mass displacement 
vector between
molecules $i$ and $j$ and $\vec{F}_{ij}$ is the total force acting
between molecules $i$ and $j$. The force between molecules is 
the sum of the atomic forces~\cite{todd:ms:2007}, i.e. 
 $\vec{F}_{ij} = \sum_{\alpha \in i} \sum_{\beta \in
  j}\vec{F}_{\alpha \beta}$, where atom $\alpha$ is a constituent atom in
molecule $i$ and $\beta$ in molecule $j$. $V$ is the system volume. 
The pressure tensor defined
above is in general not symmetric~\cite{todd:ms:2007}, and  
the traceless symmetric part 
$\stackrel{\mathrm{os}}{\tensor{P}}$ can be extracted using 
$\stackrel{\mathrm{os}}{\tensor{P}} = \frac{1}{2}(\tensor{P} + 
\tensor{P}^T) - \frac{1}{3}\mathrm{trace}(\tensor{P})$. 
From this we can define the shear stress autocorrelation function
\begin{equation}
\label{eq:sacf}
C_P(t) =  \frac{1}{3}\sum_{(\alpha \beta)}
\langle \stackrel{\mathrm{os}}{\tensor{P}}_{(\alpha \beta)}(0)
\stackrel{\mathrm{os}}{\tensor{P}}_{(\alpha \beta)}(t) \rangle \ ,
\end{equation}
where $(\alpha \beta)$ runs over the $xy, xz$, and $yz$ pressure 
tensor elements and $\langle \ldots \rangle$ denotes an ensemble
average. The ensemble average is considered equivalent to a time average.
Each sample interval is divided into sample windows for $C_p$ 
with time spans of around 2 nsec. The product
$\stackrel{\mathrm{os}}{\tensor{P}}_{(\alpha \beta)}(t_0)
\stackrel{\mathrm{os}}{\tensor{P}}_{(\alpha \beta)}(t_0+t)$ is computed
for a time $t$, $0 \geq t \geq 2$ nsec. and
for different initial times $t_0$ within the specific window. The
average is performed over the initial times within a period and over
the different periods, leading to the stress autocorrelation function
for time $t$ going from zero to 2 nsec. Comparing the
results of the 8 successive samples in a production run enables us to
probe the equilibration process. 

\begin{figure}
  \scalebox{0.275}{\includegraphics{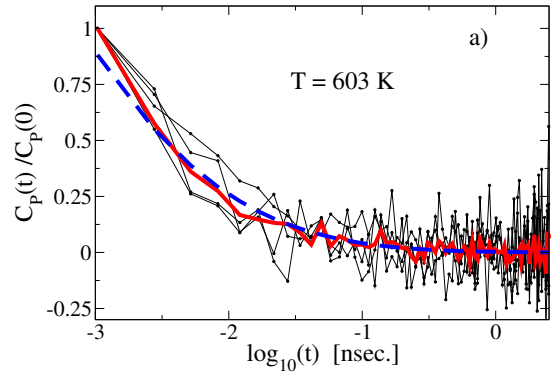}}
  \scalebox{0.275}{\includegraphics{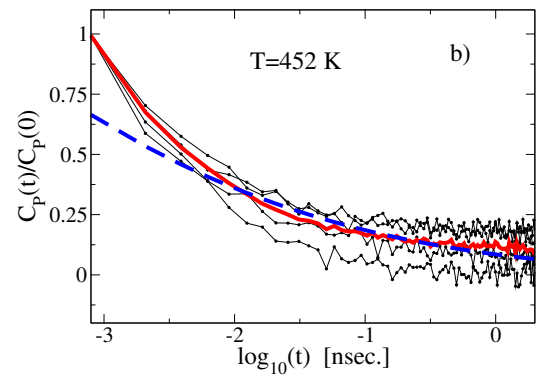}}
  \caption{\label{fig:sacf}
    (Color online). Normalized stress autocorrelation
    functions defined in Eq. (\ref{eq:sacf}) as a function of time for
    4 samples among the 10 successive ones in a production run
    (dots connected with black lines). 
    Red curves are the average of the last eight sample
    intervals, blue dashed line is the best fit of the form given in
    Eq. (\ref{eq:kww}) to the averaged data from zero to 0.63
    nsec. 
  }
\end{figure}

In Fig. \ref{fig:sacf} the stress autocorrelation function
is plotted for $T=603$ K and for $T$=452 K. 
For higher temperatures, (Fig. \ref{fig:sacf} (a)), the
correlation function is fully decayed (note the logarithmic scale of
the abscissae) and time invariant within statistical error, as shown by the
equivalent results obtained for different samples.  
For low temperatures (Fig. \ref{fig:sacf} (b)), it is seen that the
autocorrelation function does not relax fully 
and that the 8 successive samples in a production run does not lead to the same
result, i.e. the relaxation
time characterizing collective dynamics is larger than the sample
window for the stress autocorrelation function of 0.17 $\mu$sec. 
The 2 nsec. period used here is approximately ten times
longer than what was used by Z \& G to study their bitumen model
(including equilibrium properties) at temperatures 298-443
K~\cite{zhang:jcp:2007}. 

For highly viscous liquids the relaxation processes are often quantified by 
non-exponential functions~\cite{cavagna:2009}. In particular, relaxation
at later times is frequently described by a stretched exponential function 
\begin{equation}
C(t) = C(0) \exp(-(t/\tau_c)^\beta) \ , \label{eq:kww}
\end{equation}
where $\tau_c$ is the characteristic relaxation time and $\beta
<1$. Figure \ref{fig:sacf} shows the best fits of Eq. (\ref{eq:kww}) to 
the averaged data (red line). It is seen that the stretched exponential form fits data well
even for small times for $T=603$ K. For this temperature the
coefficient $\beta$ is around 0.20 which may be interpreted as an
indication of strong spatial dynamical
heterogeneity~\cite{cavagna:2009}. We discuss this point further
below.

\begin{figure}[H]
  \scalebox{0.275}{\includegraphics{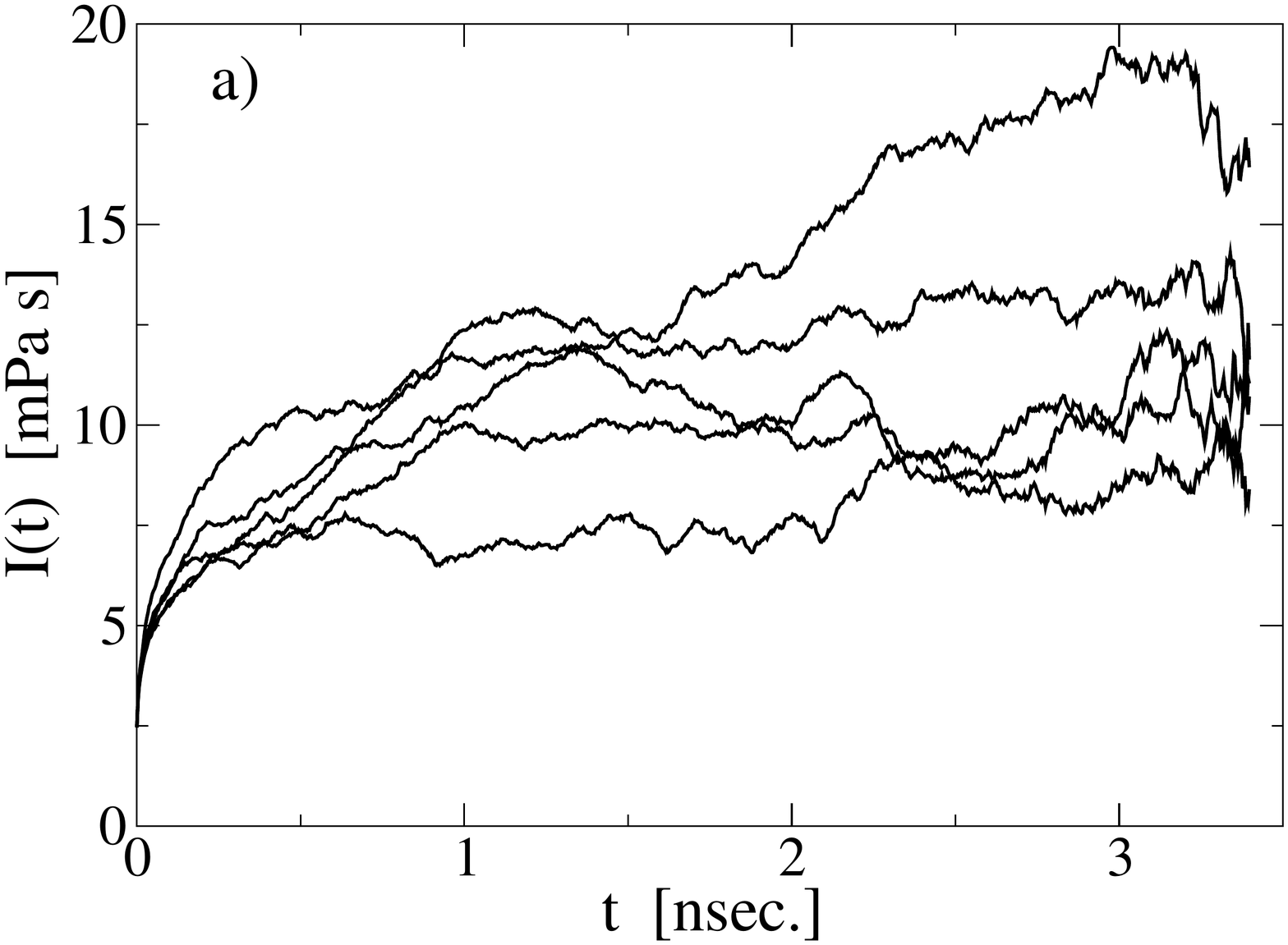}} \\
  \scalebox{0.275}{\includegraphics{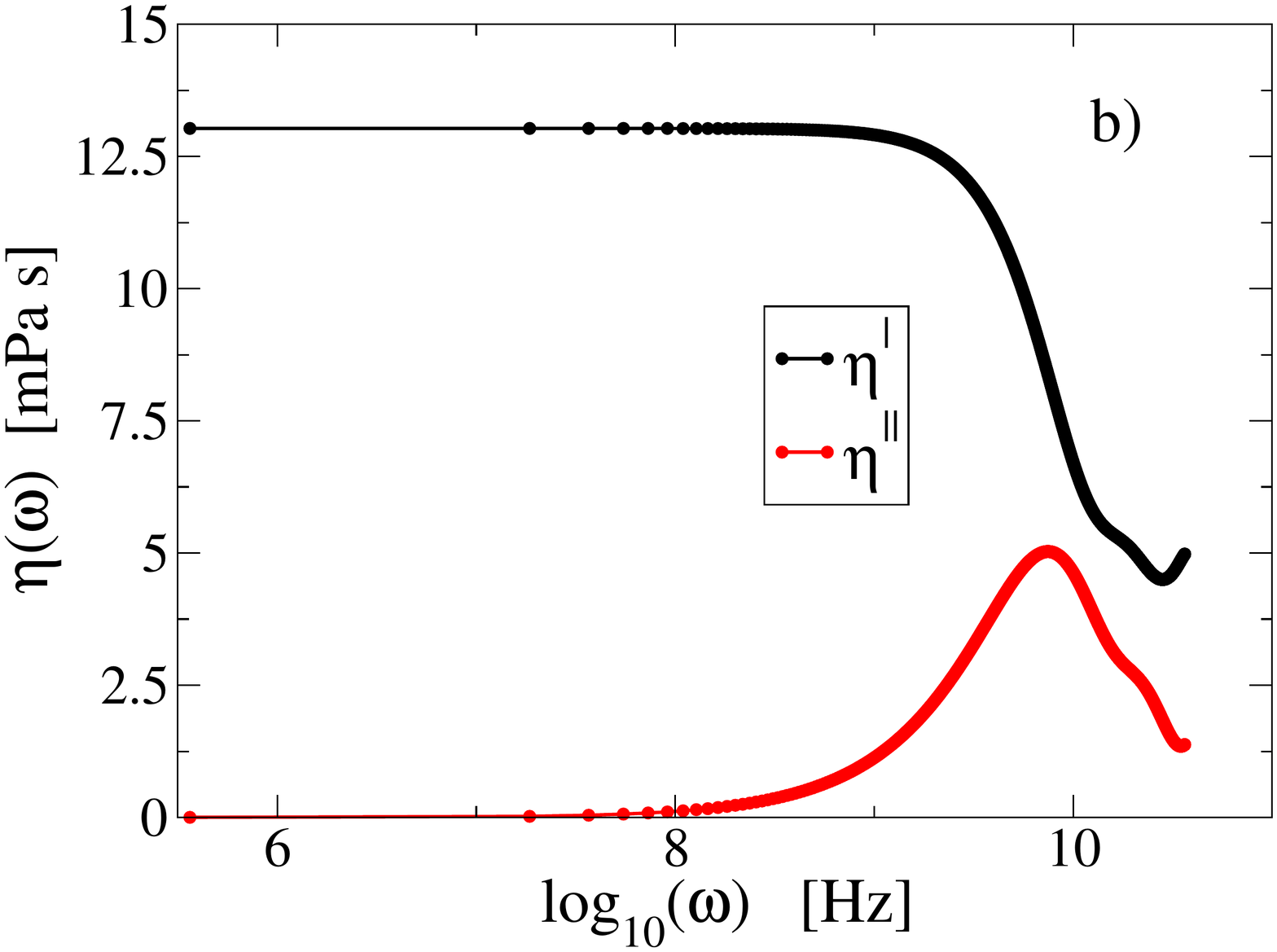}} \\
  \scalebox{0.275}{\includegraphics{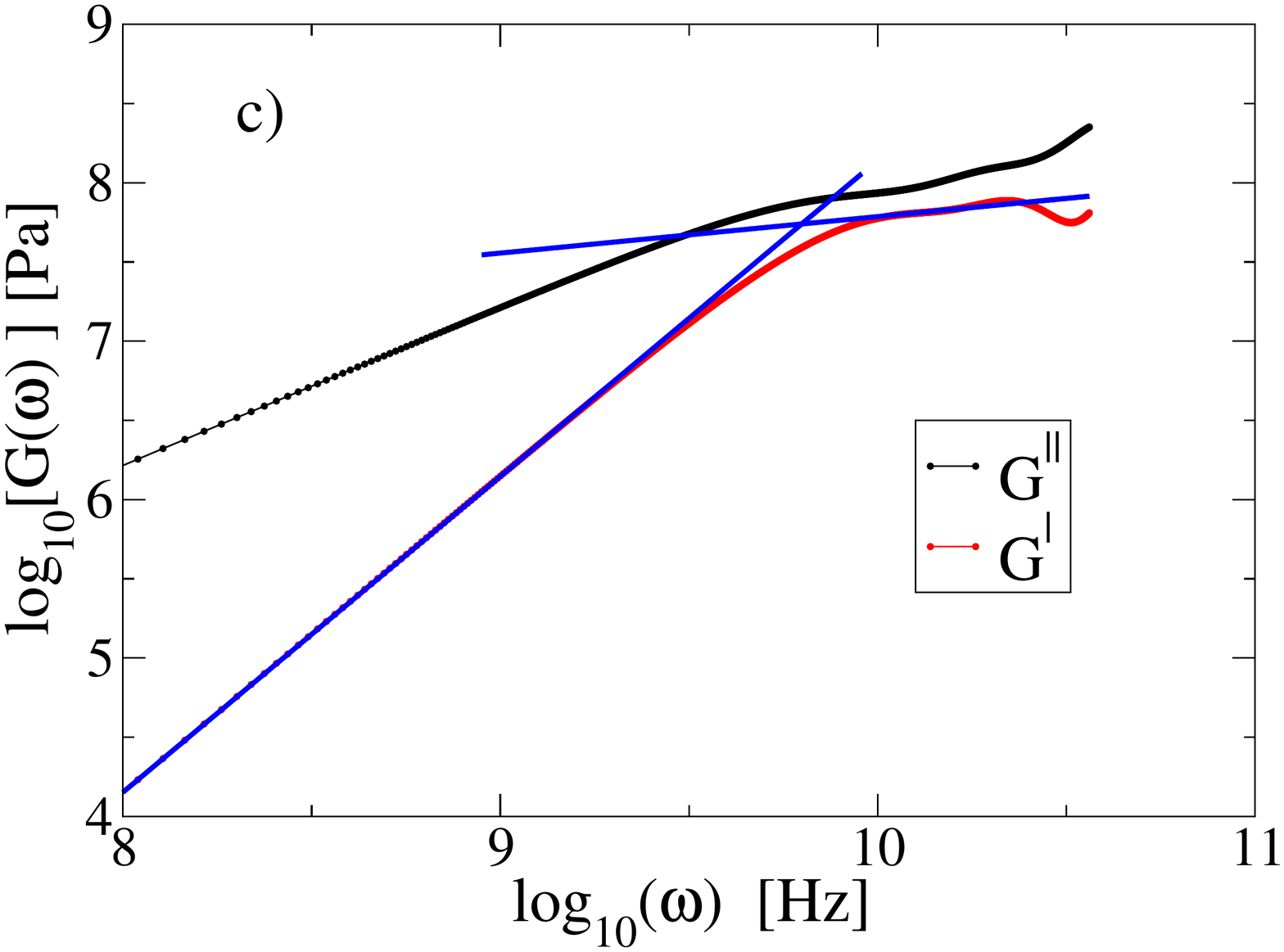}}
  \caption{\label{fig:eta_w}
    For $T$=603 K.
    a): Running integral $I(t) = \int_0^t
    C_P(s)\, ds $ of the stress autocorrelation function for
    5 samples among the 10 successive ones in a production run.
    b): Frequency-dependent viscosity as a function of
    frequency. Data was averaged and a Hann window applied to the
    single data set prior to  the numerical Fourier-Laplace
    transform.
    c): The corresponding-frequency dependent shear modulus
    $G(\omega) = G'(\omega) + iG''(\omega) = i\omega \eta(\omega)$.
    The two blues lines are fits of the form $G' \propto \omega^2$
    and $G' \propto \omega^{1/4}$.
  }
\end{figure}

The frequency-dependent viscosity $\eta(\omega) = \eta'(\omega) +
i\eta''(\omega)$ is given by the stress autocorrelation
function as follows~\cite{hansen:2006}  
\begin{eqnarray}
\eta(\omega) = \frac{V}{k_BT}\int_0^\infty 
C_P(t) \exp(- i\omega t)\, dt \ .
\label{eq:eta_w}
\end{eqnarray}
This expression can, of course, only be used for temperatures where
the stress autocorrelation function is fully relaxed and time
invariant; in Fig. \ref{fig:eta_w} (a) we have plotted the running
integral $I(t) = \int_0^t C_P(s)\, ds $ indicating that this is the case within statistical
uncertainty for $T$=603 K.

Figure \ref{fig:eta_w} (b) displays $\eta(\omega)$ for  
$T$=603 K; it is observed that the fluid response is purely viscous 
for $\omega < 100$ MHz. The imaginary part of the viscosity
exhibits a peak at around 0.5 GHz, which corresponds to a
characteristic relaxation time of approximately 0.2
nsec. The frequency-dependent shear modulus 
$G(\omega) = G'(\omega) + iG''(\omega) = i\omega \eta(\omega)$ is
plotted in Fig. \ref{fig:eta_w} (c). Two distinct regimes are
clearly observed; at low frequencies $G' \propto \omega^2$ which is the zero-frequency
behavior and at high frequencies $G' \propto
\omega^{1/4}$. This later frequency dependency is different from 
models for simple polymers~\cite{bird:1987}, so
great care should be taken when applying the theoretical
predictions from polymer science to the field of bitumen. 

\begin{figure}[H]
  \scalebox{0.275}{\includegraphics{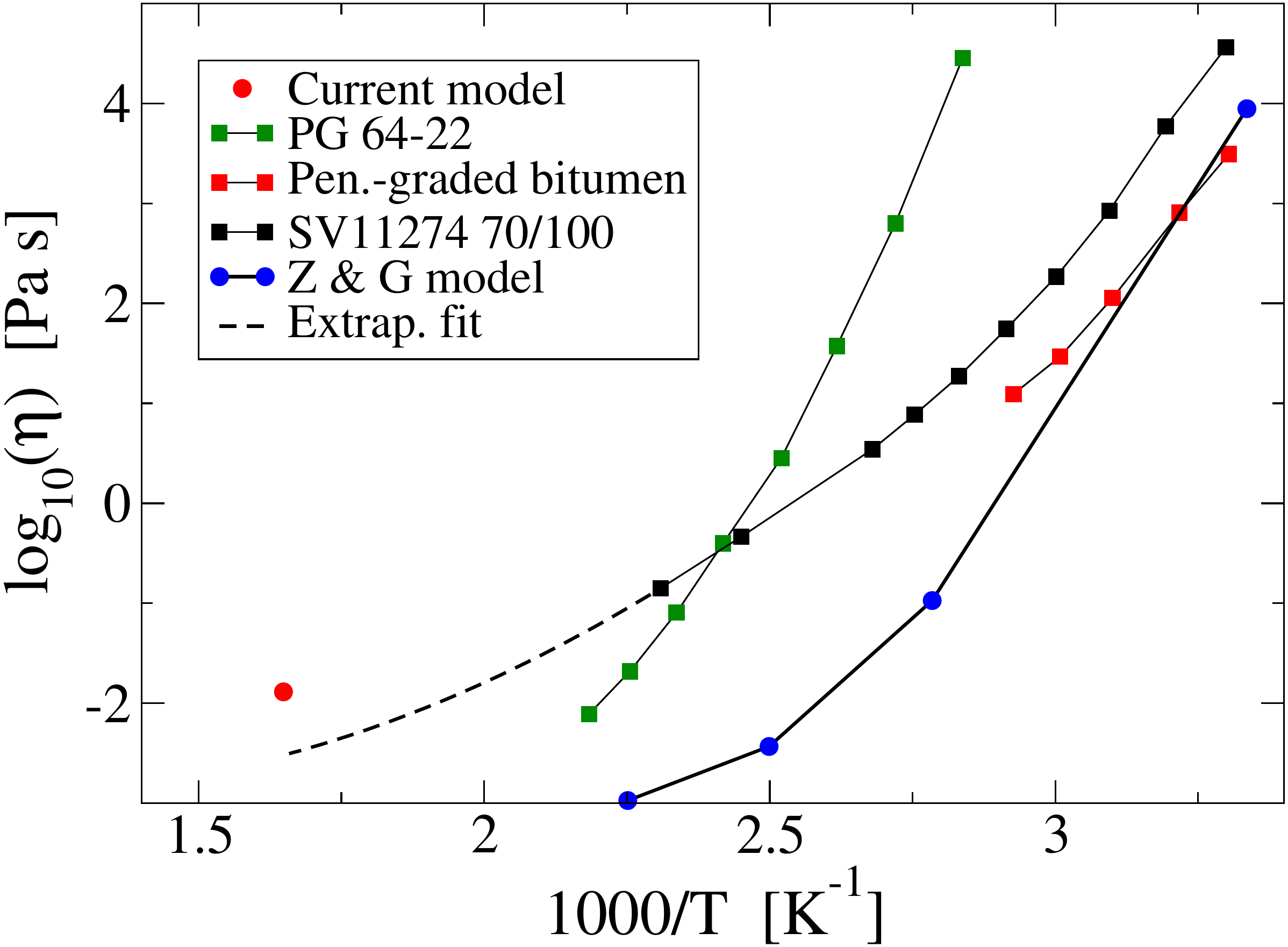}}
  \caption{\label{fig:arrh_eta}
    Comparison of the zero-frequency viscosity (red circles) with
    experimental data (filled squares)\cite{erik:SV}. The predictions from
    the Z \& G's model are also shown (blue circles connected with
    line). Data for PG 64-22 and penetration-graded bitumen is
    extracted from Ref. \onlinecite{zhang:jcp:2007}.
  }
\end{figure}

From Fig. \ref{fig:eta_w} (b) we can extract the zero-frequency
viscosity for $T$ = 603 K. In order to compare the predicted viscosity
with experimental data available for lower temperatures we
extrapolate these to $T$ = 603 K. In Fig.~\ref{fig:arrh_eta} this
is done for the standard SV11274-70/100 bitumen, which is 
measured in this work, see Ref. \onlinecite{erik:SV} for a short
description of the measurement technique. Only data for the three 
highest temperatures have been used for the
fitting procedures.  Data were fitted to (i) first order polynomial
(pure Arrhenius behavior) (ii) the Vogel-Fulcher-Tamman (VFT) function
$\eta_0 \propto \exp(\alpha/(T-T_0)$ where $\alpha$ and $T_0$ are fitting
parameters, and (iii) a power law (shown). The power
law is found to fit data best.
The model predicts the viscosity 
reasonably well for SV11274-70/100 bitumen; but not for PG 64-22
bitumen. As we mentioned above, it is not the purpose of the model to
fit a specific bitumen, and the fact that the model
cannot capture the dynamical properties of different bitumens is expected.
Note that we used the stress autocorrelation function
to calculate the viscosity, which is only possible for high
temperatures. Other approximative
methods~\cite{bird:1987,zhang:jcp:2007} may be used to estimate the
viscosity; however it is not clear whether these methods can be 
applied to bitumen mixtures.

\subsection{Rotational dynamics}
Above it was observed that while the molecular mean-square
displacements indicate a well-defined diffusive regime for all
temperatures studied, the stress autocorrelation functions 
showed a not fully relaxed system for times around 0.17 $\mu$sec. 
To better understand this behavior of the model we will study the
rotational dynamics for resinous oil and asphaltene.
Rotation is quantified through the dynamics of the normalized
orientational vector $\vec{u}(t)$ normal to one of the ring
structures. As in Sect. \ref{sec:model}, the vector is defined as the
cross product of two vectors pointing along a chemical bond as
illustrated in Fig. \ref{fig:molecules}, where the numbers indicate
the atom indices involved in the definition of $\vec{u}$. 

\begin{figure}[H]
  \scalebox{0.27}{\includegraphics{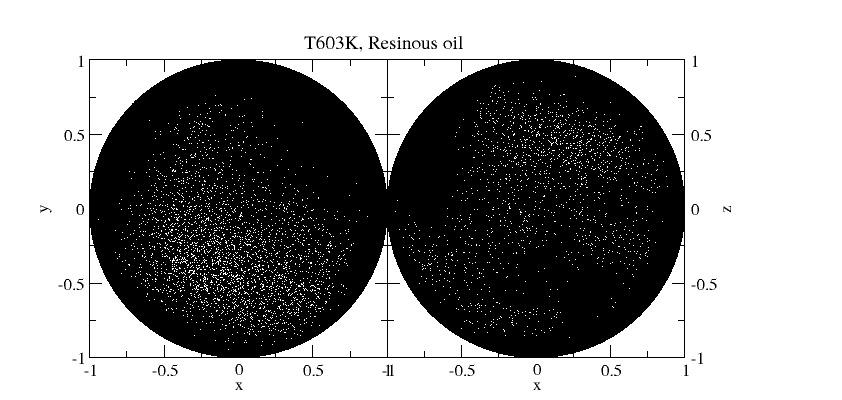}}
  \scalebox{0.27}{\includegraphics{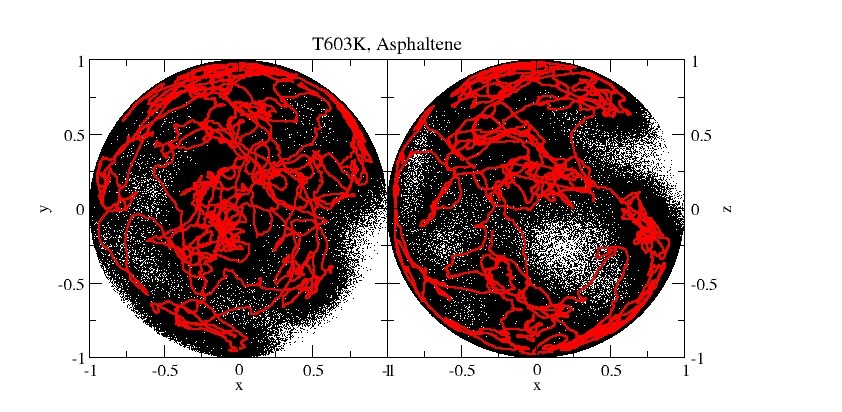}} \\
  \scalebox{0.27}{\includegraphics{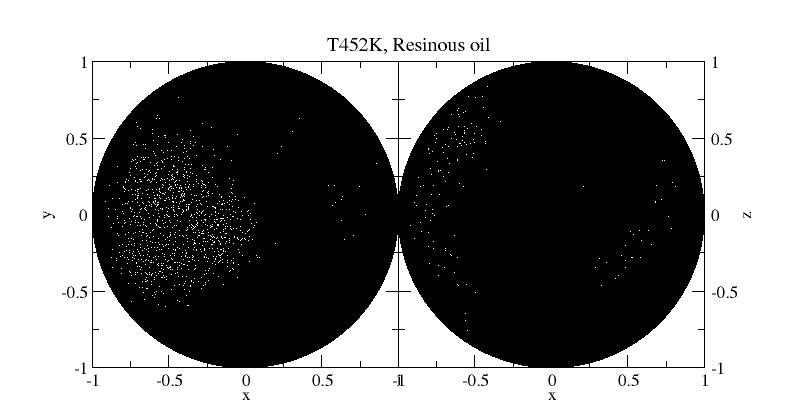}}
  \scalebox{0.27}{\includegraphics{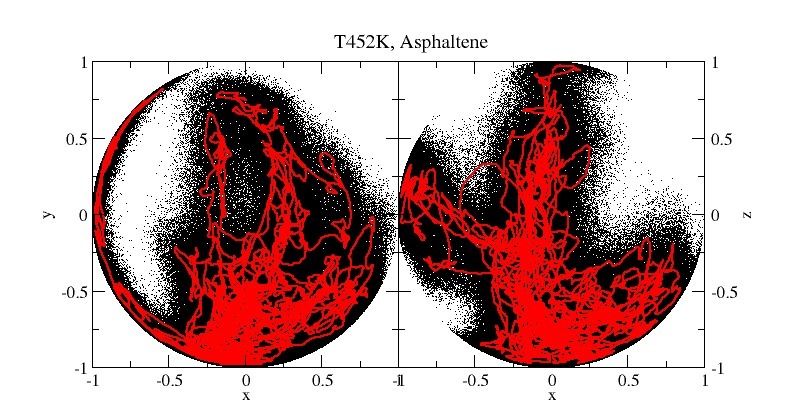}}
  \caption{\label{fig:orient}
    (Color online). Projections of the normalized orientation vector
    $\vec{u}$ as a function of time. Each dot represents the end point
    of the vector. For the asphaltenes: red lines are averages of 1000
    data points.
  }
\end{figure}

In Fig. \ref{fig:orient}, projections of the
orientational vector are plotted as it evolves over time.  
It is evident that the smaller resinous oil goes through all
possible orientations (to a good approximation) for all temperatures
within the 0.85 to 1.75 $\mu$sec. sample time. This is
not the case for the larger asphaltene. Especially for lower
temperatures the orientation of this molecule is confined to a
small subset of orientations. Naturally, this subset increases
as a function of time.  

\begin{figure}[H]
  \scalebox{0.3}{\includegraphics{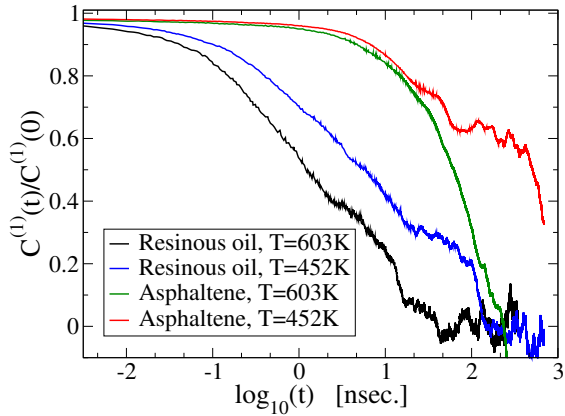}}
  \caption{\label{fig:C_1}
    (Color online). Rotational correlation function $C^{(1)}$ for
    small times. 
  }
\end{figure}
The first-order rotational correlation function,
$C^{(1)}(t) =
\langle\vec{u}(0)\cdot\vec{u}(t)\rangle$~\cite{hansen:2006},
is plotted in Fig. \ref{fig:C_1}.
The rotational relaxation appears to be governed by multiple
relaxation times because of the nonsymmetric molecular structures.
From $C^{(1)}(t)$ it is possible to define a
characteristic relaxation time $\tau_r = \int_0^\infty C_r(t) \,
dt/C_r(0)$. For example, we obtain $\tau_r=5 $ nsec. for the
resinous oil at 452 K. Again this is evidence of a much slower
dynamics compared to Z \& G's model where $\tau_r = 1.44$
nsec. for the asphaltene at 443.15 K.  
In combination with Fig. \ref{fig:msd} it can be
concluded that while, on average, the asphaltene molecules enter the diffusive regime,
the single molecules keep their orientation relatively fixed. Such a decoupling
between the rotational and diffusional dynamics is known from many experiments
on viscous liquids, see Ref. \onlinecite{edinger:jcp:2012} and references therein.

\begin{figure}[H]
  \scalebox{0.6}{\includegraphics{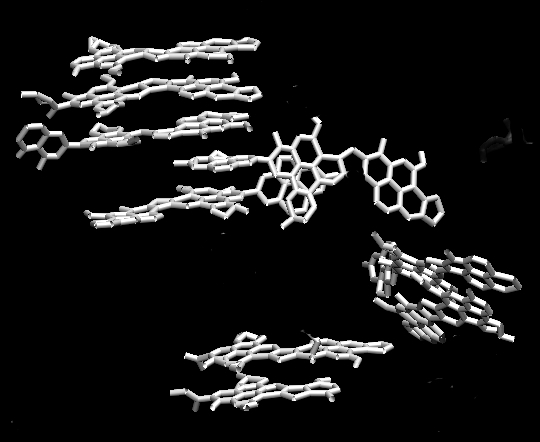}}
  \caption{\label{fig:snap}
    Snapshot of the configuration of the asphaltenes. Nine molecules
    are shown, and parts of the molecules has been removed for
    clarity.
  }
\end{figure}

The snapshot shown in Fig. \ref{fig:snap} highlights this
picture; the asphaltene molecules tend to align and form
nano-aggregates, but occasionally they leave the 
aggregate and then later attach at a new position or aggregate. This dynamical
behavior reveals that the diffusivity of the asphaltenes is determined by
the mobility of just a small fraction of the molecules at any instant in
time. The same remark applies to resin and resinous oil that also
participate in the nano-aggregate formation. However, resin and
resinous oil evolve at a  different time scale.
This long-time fixed orientation and clustering naturally leads to
a very slowly decaying stress autocorrelation function. We conclude
that the dynamics of the model is dominated by dynamical
heterogeneities. 

\section{Summary}
\label{sec:summary}
In this paper we have proposed a new molecular model of bitumen. 
Compared to Z \& G's model, we have replaced the dimethylnaphtalene
with two higher-weight molecular
structures representing resin and resinous oil molecules. The new model is a
united-atomic-unit model, which suppresses the high-frequency modes and
enables long simulation times. Using GPU-based software we are
able to simulate time spans in order of $\mu$sec., rather than
nsec.~\cite{zhang:ef:2007:2}. This makes it possible to study
additional slow relaxation processes characterizing bitumen.  

For the density and temperatures studied here
the model is characterized by slow dynamics, 
which changes little over a relatively large temperature range.
The model predictions are in reasonable agreement with
experimental data for diffusion and viscosity. Better
agreement with specific bitumen solutions can be obtained by model 
re-parametrization and by choosing the state point carefully. 

The model captures the heterogeneous dynamics since the asphaltene, resin
and resinous oil form nano-aggregates. Murgich et
al.~\cite{murgich:ef:1996} and Z \& G~\cite{zhang:ef:2007:2} have
reported that the asphaltene molecules tend to align, which
supports this picture. The fact that asphaltenes from nano-aggregates
is also known from experimental work~\cite{yen:ac:1961,mullins:ef:2012}
The characteristic dynamical relaxation time of
these localized aggregates are different from the more homogeneously 
distributed parts of the system leading to a high degree of dynamical
heterogeneity. This is also evident in the stretched exponential fit of
the stress autocorrelation function. 
In Fig. \ref{fig:arrh_diff} one observes separate time scales, 
the slowest one being due to the dynamics of larger asphaltene
molecules. This indicates that the asphaltene nano-aggregate dynamics may be
studied in more details on a longer time scale using coarse graining
techniques like Brownian dynamics. This may form the foundation for
understanding the dynamical heterogeneity of bitumen. 

\acknowledgements
This work is sponsored by the Danish Council for Strategic Research as
part of the Cooee project.  The centre for viscous liquid dynamics
``Glass and Time'' is sponsored by the Danish National Research
Foundation (DNRF).

\bibliographystyle{unsrt}
\bibliography{../lit_bitumen.bib}

\end{document}